# No More Perfect Codes: Classification of Perfect Quantum Codes

Zhuo Li and Lijuan Xing

*Abstract*—We solve the problem of the classification of perfect quantum codes. We prove that the only nontrivial perfect quantum codes are those with the parameters $(((q^{2l}-1)/(q^2-1), q^{n-2l}, 3))_q$. There exist no other nontrivial perfect quantum codes.

*Index Terms*—Perfect quantum codes, quantum error-correcting codes, quantum information.

## I. INTRODUCTION

Quantum information can be protected by encoding it into a quantum error-correcting code [1]-[7]. For any prime power $q$, a $q$-ary quantum code with parameters $((n, K, d))_q$ is a $K$-dimensional subspace of the state space $\mathcal{H} = (\mathbb{C}^q)^{\otimes n}$ of $n$ quantum systems with $q$ levels that can detect all errors affecting less than $d$ quantum systems, but cannot detect some errors affecting $d$ quantum systems. In the construction of quantum codes, one would like to have both large dimension $K$ and large minimum distance $d$, but these are two conflicting requirements on the quantum code. The trade off between the number of correctable errors and the size of the quantum code is usually quantified by various bounds. For example, a pure $((n, K, d))_q$ quantum code satisfies the quantum Hamming bound

$$K \sum_{i=0}^{\lfloor \frac{d-1}{2} \rfloor} (q^2-1)^i \binom{n}{i} \leq q^n. \quad (1)$$

A pure quantum code for which equality holds in (1) is called perfect. Just as the name suggests, a perfect quantum code possesses perfect structure and properties that greatly interests people. So finding perfect quantum codes has been one of the research focuses in the field of quantum coding.

Two types of perfect quantum codes were discovered in the early 2000's:

(i) The cyclic Hamming codes $((n, q^{n-2m}, 3))_q$ with $m \geq 2$, $\gcd(m, q^2-1) = 1$, $n = (q^{2m}-1)/(q^2-1)$, see [8]-[10].

Zhuo Li and Lijuan Xing are with the State Key Laboratory of Integrated Service Networks, Xidian University, Xi'an, Shannxi 710071, China (e-mail: lizhuo@xidian.edu.cn).

(ii) The perfect twisted codes $((n, q^{n-r-2}, 3))_q$ with $n = (q^{r+2}-1)/(q^2-1)$, $r \geq 2$, $r$ even, see [11].

Finally there is the trivial perfect quantum code: a code containing the whole space. Subsequently people wondered if other perfect quantum codes could be discovered, or no others existed. To this date this problem remains open.

In this correspondence, we solve this problem completely. We prove that the only nontrivial perfect quantum codes are those with the parameters of the quantum Hamming or twisted codes. So, in the future researches on perfect quantum codes, one can pay attention only to these codes, not to finding others. Our approach is based on the group algebra framework of quantum codes. In the following we shall use the language of this framework, for the details of which one can see [12].

## II. CLASSIFICATION

**Lemma 1.** Let $Y_i = \sum_{\mathrm{wt}(v)=i} z^v \in \mathbb{C}Z$. If $u \in G^n$ has weight $w$,

$$\chi_u(Y_i) = \sum_{r=0}^{i} (-1)^r (q^2-1)^{i-r} \binom{n-w}{i-r} \binom{w}{r} = P_i(w),$$

where $P_i(x)$ is a Krawtchouk polynomial.

*Proof.* From the definition of the character,

$$\chi_u(Y_i) = \sum_{\mathrm{wt}(v)=i} \prod_{j=1}^{n} \omega_{u_j v_j} \overline{\omega}_{v_j u_j}.$$

There are $(q^2-1)^{i-r} \binom{n-w}{i-r} \binom{w}{r}$ vectors $v$ of weight $i$ which have $i-r$ nonzero components in the $n-w$ coordinates where $u$ is 0, and $r$ nonzero components in the $w$ coordinates where $u$ is nonzero. Each of these vectors $v$ contributes $(-1)^r$ to the sum.

Let $\mathcal{C}$ be an $((n, K, d))_q$ quantum code with the orthogonal projector $P$, and let $\{A_i\}$ and $\{A'_i\}$ be the Hamming weight distributions and the dual Hamming weight distributions of $\mathcal{C}$ respectively. The annihilator polynomial of $\mathcal{C}$ is defined to be

$$\alpha(x) = \frac{q^n}{K} \prod_{j=1}^{s} \left(1 - \frac{x}{\sigma_j}\right),$$

where $0, \sigma_1, \sigma_2, \ldots, \sigma_s$ are the subscripts $i$ for which $A_i \neq 0$. Note that for $0 < i \leq n$ either $\alpha(i) = 0$ or $A_i = 0$.

The expansion of $\alpha(x)$ in terms of Krawtchouk polynomials,

$$\alpha(x) = \sum_{i=0}^{s} \alpha_i P_i(x),$$

is called the Krawtchouk expansion of $\alpha(x)$, and the $\alpha_i$ are called the Krawtchouk coefficients. The expansion stops at $P_s(x)$ since $\alpha(x)$ is of degree $s$. Also $\alpha_s \neq 0$ and

$$\alpha_i = \frac{1}{q^{2n}} \sum_{k=0}^{n} \alpha(k) P_k(i).$$

**Theorem 2.**

$$\sum_{i=0}^{s} \alpha_i \sum_{\mathrm{wt}(g)=i} E_g P E_g^\dagger = I.$$

*Proof.* Let $a_h = \mathrm{tr}\, E_h^\dagger P / q^n$. Then

$$\mathrm{LHS} = \sum_{i=0}^{s} \alpha_i \sum_{\mathrm{wt}(g)=i} E_g P E_g^\dagger = \sum_{i=0}^{s} \alpha_i \sum_{\mathrm{wt}(g)=i} E_g \left(\sum_{h \in G^n} a_h E_h\right) E_g^\dagger$$

$$= \sum_{i=0}^{s} \alpha_i \sum_{h \in G^n} a_h \overline{\chi}_h(Y_i) E_h = \sum_{h \in G^n} a_h E_h \sum_{i=0}^{s} \alpha_i \overline{\chi}_h(Y_i).$$

First, if $h$ has weight $w > 0$, by Lemma 1

$$\sum_{i=0}^{s} \alpha_i \overline{\chi}_h(Y_i) = \sum_{i=0}^{s} \alpha_i \overline{P_i}(w) = \sum_{i=0}^{s} \alpha_i P_i(w) = \alpha(w).$$

If $w$ is one of the $\sigma_j$'s, $\alpha(w) = 0$ by definition, but if not, $a_h = 0$ since $A_w = 0$. Thus in either case

$$a_h E_h \sum_{i=0}^{s} \alpha_i \overline{\chi}_h(Y_i) = 0, \quad h \neq 0.$$

Second, if $h = 0$, $a_0 E_0 \alpha(0) = (K/q^n) I (q^n/K) = I$.

**Lemma 3.** If

$$\beta(x) = \sum_{i=0}^{t} \beta_i P_i(x)$$

has the property that

$$\sum_{i=0}^{t} \beta_i \sum_{\mathrm{wt}(g)=i} E_g P E_g^\dagger = I,$$

then the annihilator polynomial $\alpha(x)$ of $\mathcal{C}$ divides $\beta(x)$.

*Proof.* From the proof of Theorem 2, it follows that all the $\sigma_j$'s must be the zeros of $\beta(x)$.

In the following we shall prove that a pure $((n, K, d = 2e+1))_q$ quantum code is perfect if and only if $s = e$. We conclude with an important necessary condition for a code to be perfect.

**Theorem 4.** For any pure quantum code, $s \geq \frac{1}{2}(d-1)$.

*Proof.* First suppose $d$ is odd and let $h$ be a vector of weight $\frac{1}{2}(d-1)$. If $s < \frac{1}{2}(d-1)$ then by the definition of pure quantum codes

$$\left(\sum_{i=0}^{s} \alpha_i \sum_{\mathrm{wt}(g)=i} E_g P E_g^\dagger\right) E_h P = \sum_{i=0}^{s} \alpha_i \sum_{\mathrm{wt}(g)=i} E_g (P E_g^\dagger E_h P) = 0,$$

a contradiction with Theorem 2. Similarly if $d$ is even.

**Lemma 5.** For any pure quantum code, if $s = \frac{1}{2}(d-1)$ ($d$ is odd) then $\alpha(x) = P_0(x) + P_1(x) + \cdots + P_{\frac{1}{2}(d-1)}(x)$.

*Proof.* Let $h$ be a vector of weight $w \leq s$. Then by Theorem 2,

$$E_h P = \left(\sum_{i=0}^{s} \alpha_i \sum_{\mathrm{wt}(g)=i} E_g P E_g^\dagger\right) E_h P$$

$$= \sum_{i=0}^{s} \alpha_i \sum_{\mathrm{wt}(g)=i} E_g (P E_g^\dagger E_h P) = \alpha_w E_h P,$$

so $\alpha_w = 1$ for $w = 0, 1, \ldots, s$.

**Theorem 6.** A pure $((n, K, d = 2e+1))_q$ quantum code $\mathcal{C}$ is perfect iff $s = e$.

*Proof.* Suppose $s = e$. Then by Theorem 2 and Lemma 5

$$\sum_{i=0}^{e} \sum_{\mathrm{wt}(g)=i} E_g P E_g^\dagger = I, \qquad (2)$$

which says that $\mathcal{C}$ is perfect. Conversely, suppose $\mathcal{C}$ is perfect, so that (2) holds.

We apply Lemma 3 with $\beta(x) = P_0(x) + \cdots + P_e(x)$ and deduce that the annihilator polynomial $\alpha(x)$ of $\mathcal{C}$ must divide $\beta(x)$. Hence

$$e = \deg \beta(x) \geq \deg \alpha(x) = s.$$

But $s \geq e$ by Theorem 4.

Thus the annihilator polynomial of a perfect quantum code is

$$\alpha(x) = P_0(x) + \cdots + P_e(x).$$

This is called Lloyd's polynomial and is denoted by $L_e(x)$. It follows that:

**Theorem 7.** If there exists an $((n, K, 2e+1))_q$ perfect quantum code, then the Lloyd polynomial

$$L_e(x) = P_0(x;n) + \cdots + P_e(x;n) = P_e(x-1;n-1)$$
$$= \sum_{j=0}^{e} (-1)^j (q^2-1)^{e-j} \binom{x-1}{j}\binom{n-x}{e-j}$$

has $e$ integer zeros $\sigma_1, \ldots, \sigma_e$ satisfying $0 < \sigma_1 < \cdots < \sigma_e < n$.

Let $\mathcal{C}$ be an $((n, K, 2e+1))_q$ perfect quantum code, where $q = p^r$, $p$ is a prime. Then the following relations hold.

**Lemma 8.** The dimension $K$ is a power of $q$, and

$$\sum_{i=0}^{e} (q^2-1)^i \binom{n}{i} = q^{2l}$$

for some integer $l$.

*Proof.* Since the code is perfect,

$$K \sum_{i=0}^{e} \binom{n}{i}(q^2-1)^i = q^n = p^{nr}.$$

Therefore $K = p^j$ and

$$\sum_{i=0}^{e} \binom{n}{i}(q^2-1)^i = p^{nr-j}.$$

Thus $q^2 - 1 = p^{2r} - 1$ divides $p^{nr-j} - 1$. Hence $2r$ divides $nr - j$, $r$ divides $j$ and $K$ is a power of $q$.

By Theorem 7 and Lemma 8 we can obtain the key result of this correspondence:

**Theorem 9.** For any prime power $q$, a nontrivial $q$-ary perfect quantum code must have the parameters

$$\left(\left(n = \frac{q^{2l}-1}{q^2-1}, K = q^{n-2l}, d = 3\right)\right)_q.$$

*Remark.* The proof of this theorem is long, and exactly follows that of the Tietavainen-Van Lint theorem in the classical coding theory [13, Ch. 6, Theorem 33]. Thus the proof is omitted here.

So far we have solved the problem of the classification of perfect quantum codes. Theorem 9 tells us that besides the trivial one of the whole space, the only perfect quantum codes are those with the parameters $\left(\left((q^{2l}-1)/(q^2-1), q^{n-2l}, 3\right)\right)_q$. For example, the Hamming codes and the twisted codes mentioned above are such codes. No other perfect quantum codes do exist.